\documentclass[draft,preprint,aps,
showpacs
]{revtex4}

\usepackage{epsfig}

\usepackage{graphicx}
\usepackage{bm}

\input{epsf}

\newcommand{\be}{\begin{equation}}
\newcommand{\ee}{\end{equation}}
\newcommand{\bea}{\begin{eqnarray}}
\newcommand{\eea}{\end{eqnarray}}
\newcommand{\nn}{\nonumber \\}

\newcommand{\Tr}{\mbox{Tr}}
\newcommand{\tr}{\mbox{tr}}

\begin{document}

\preprint{Guchi-TP-023}
\date{\today%
}
\title{Divergences in QED on a Graph}

\author{Nahomi Kan}
\email{b1834@sty.cc.yamaguchi-u.ac.jp}
\affiliation{Graduate School of Science and Engineering, Yamaguchi University, 
Yoshida, Yamaguchi-shi, Yamaguchi 753-8512, Japan}

\author{Kiyoshi Shiraishi}
\email{shiraish@yamaguchi-u.ac.jp}
\affiliation{Faculty of Science, Yamaguchi University,
Yoshida, Yamaguchi-shi, Yamaguchi 753-8512, Japan}

\begin{abstract}
We consider a model of quantum electrodynamics (QED) on a graph. 
The one-loop divergences in the model are investigated by use of the background
field method.
\end{abstract}



\maketitle


\section{Introduction}
One of the most important problems in modern particle physics is to understand
the existence of largely
different mass scales within a unified theory.
In many occasions, we use the idea of supersymmetry (SUSY) to control the quantum
correction in order to understand the mass hierarchy. 

Recently, the scenario for the electroweak symmetry breaking without SUSY
is suggested and studied by many authors.%
\footnote{The seminal paper in recent development is \cite{ACGE}.}
The divergent diagrams are 
mutually cancelled among contributions from a number of bosonic fields.

The basic idea of such a mechanism is now attributed to dimensional
deconstruction~\cite{ACG}, where copies of a
four-dimensional `theory' as well as a new set of fields linking pairs of these 
`theories' are considered. 
Then the resulting whole theory given by the `theory space' may be equivalent to a 
higher-dimensional theory with discretized extra dimensions.

The present authors considered previously a generalization of the
deconstruction~\cite{KSP}. \footnote{A generalization of deconstruction is also argued in ref.~\cite{NOS}.} 
We identify the theory space as a graph consisting of
vertices and edges. In the present paper, we further investigate the divergences in
the field theory on a graph, particularly focusing on an Abelian theory.
Although the non-Abelian structure and alignment of fields in a certain
representation may be essential for realistic models, the substantial behavior of
divergences can be viewed from a simpler model. Recently, a model with extra
massive vector boson is studied by K\"ors and Nath~\cite{nath}, where the Stueckelberg
formalism is utilized. Our model can be applied to a generalization of their work. 

We organize the present paper in the following way.
In Sec.~\ref{sec:2}, we review graph theory and matrices associated with a graph,
including the Laplacian of a graph.
The lagrangian for gauge fields on a graph is described in Sec.~\ref{sec:3}.
The lagrangian for fermion fields is described in Sec.~\ref{sec:4}.
In Sec.~\ref{sec:5}, the one-loop logarithmical divergences in effective lagrangian
density is discussed.  In Sec.~\ref{sec:6}, we study the one-loop finiteness
of the effective potential for a constant background link scalar fields. 
The comment on the non-Abelian generalization is given in Sec.~\ref{sec:7}.
We close
with Sec.~\ref{sec:f}, where summary and prospects are given.

\section{Graph theory and matrices associated with a graph}
\label{sec:2}

In this section, after a brief description of `graph'~\cite{graphtheory},
some matrices associated with a graph are introduced.

Let $G(V,E)$ be a 
\underline{graph}
 with 
\underline{vertex} set $V$ and 
\underline{edge} set $E$.
The set of edges connects the vertices.
A graph which does not have 
\underline{multiple edges}%
~\footnote{Edges connecting the same two vertices.}
 and 
\underline{self-loops}%
~\footnote{Edges from a vertex to itself.}
is called as a 
\underline{simple graph}.
We only consider simple graphs in the present paper.
The 
\underline{order}
 of $G$, denoted by $p$ in this paper, is the number of vertices in
the graph, while the 
\underline{size} of $G$, denoted by $q$ in this paper, is the number of edges
in the graph.  
A pair of vertices $u$ and $v$ are said to be 
\underline{adjacent},
denoted $u\sim v$,
if there exists an edge $e\in E$ which connects $u$ and $v$.
Such edge is denoted as $e=\{u,v\}$.

The 
\underline{adjacency matrix} $A$ is defined as
\be
(A)_{vv'}=\left\{
\begin{array}{ccl}
1&~&{\rm if}~v\sim v'~(v~{\rm is~adjacent~to}~v')\\
0&~&{\rm otherwise}
\end{array}
\right.\, .
\ee

The 
\underline{degree} of a vertex $v$, denoted $deg(v)$, is the number of edges directly
connected to (in other words, is incident with)
$v$. The (diagonal) 
\underline{degree matrix} $D$ is defined as
\be
(D)_{vv'}=\left\{
\begin{array}{ccl}
deg(v)&~&{\rm if}~v=v'\\
0&~&{\rm otherwise}
\end{array}
\right.\, .
\ee

The 
\underline{graph Laplacian} (or 
\underline{combinatorial Laplacian}) $\Delta(G)$ is
defined~\cite{Mohar} by
\be
\Delta_{vv'}=(D-A)_{vv'}=
\left\{\begin{array}{cl}
deg(v) & ~{\rm if}~v=v'\\
-1 & ~{\rm if}~v~{\rm and}~v'~{\rm are~adjacent}~(v\sim v')\\
0 & ~{\rm otherwise}
\end{array}\right.
\, ,
\ee
where $v, v' \in V$ and $deg(v)$ denotes the degree of $v$.

The $(mass)^2$ matrix for vector fields in the Hill-Leibovich
model~\cite{HL} is proportional to
$\Delta$ for a cycle graph~\cite{graphtheory} with $N$ vertices (denoted as $C_N$).

Next we consider a 
\underline{directed graph}. An oriented edge $e=[u,v]$ ($u,v\in V(G)$)
connects the 
\underline{origin}
$u=o(e)$ and the 
\underline{terminus}
 $v=t(e)$ (and an unoriented edge does not distinguish
its origin and terminus).

The 
\underline{incidence matrix} (for a directed graph) $E$ is defined by
\be
(E)_{ve}=
\left\{\begin{array}{rcl}
1 &~& ~{\rm if}~v=o(e) \\
-1 &~& ~{\rm if}~v=t(e) \\
0 &~& ~{\rm otherwise}
\end{array}\right.
\, .
\ee

Our important observation is
\be
\Delta=EE^{T}\, ,
\ee
for any given graph.

\section{Vector fields (+scalar fields)}
\label{sec:3}

The simplest model with the Abelian symmetry is studied by Hill and
Leibovich~\cite{HL}. We consider here the extension of the model
constructed on a general graph.

We associate vector fields with vertices of a graph $G$.
Further we introduce a link field $U_e$ 
on each edge.
We can write the lagrangian density for vector fields whose (mass)${}^2$ matrix 
is $\Delta(G)$ as~%
\footnote{In this section the metric signature is $(+---)$.}
\be
{\cal
L}_V=-\frac{1}{4 g^2}\sum_{v\in V}
F^{\mu\nu}_vF_{v\,\mu\nu}+\frac{1}{2}v^2\sum_{e\in E} |D^\mu
U_e|^2\, ,
\label{eqL}
\ee
where $g$ is a gauge coupling and
$F_{v}^{\mu\nu}=\partial^{\mu}{A}_{v}^{\nu}-
\partial^{\nu}{A}_{v}^{\mu}$
($\mu, \nu=0, 1, 2, 3$) stands for the field strength.
The covariant derivative is defined as
$D^\mu U_e=\partial^\mu U_e-i(A^\mu_{o(e)}U_e-U_eA^\mu_{t(e)})$.

The lagrangian is invariant under the following gauge transformations:
\bea
A^{\mu}_{v}&\rightarrow& A^{\mu}_{v}+\partial^{\mu}\omega_v \, ,\nn
U_{e}&\rightarrow& \exp({i\omega_{o(e)}})U_{e}\exp({-i\omega_{t(e)}})\, .
\label{ga1}
\eea

Now we assume that the absolute value of each link field $|U_{k}|$ has a common
value $1$. 
If we express $U_e$ as $\exp(-i\chi_e)$, the lagrangian (\ref{eqL}) becomes
\bea
{\cal
L}_V&=&-\frac{1}{4 g^2}\sum_{v\in V}
F^{\mu\nu}_vF_{v\,\mu\nu}+\frac{1}{2}v^2\sum_{e\in E} (\partial^\mu
\chi_e+A^{\mu}_{o(e)}-A^{\mu}_{t(e)})^2\nn
&=&-\frac{1}{4 g^2}\sum_{v\in V}
F^{\mu\nu}_vF_{v\,\mu\nu}+\frac{1}{2}v^2\sum_{e\in E} (\partial^\mu
\chi_e+(E^TA^{\mu})_e)^2\, ,
\eea
where $(E^TA^{\mu})_e$ is an abbreviation form of 
$\sum_{v\in V}(E^T)_{ev}A^{\mu}_v$.

We find that the term including $\chi_e$ resembles the gauge kinetic term of the
extra ``fifth'' index, $\propto (F_{\mu 5})^2$, when we regard
$gv(A^{\mu}_{t(e)}-A^{\mu}_{o(e)})$ as a discretization of
differentiation and $A_5\propto \chi$. 
Then the gauge transformation on $\chi_e$ is $\chi_e\rightarrow\chi_e+\delta\chi_e$
with
\be
\delta\chi_e=\omega_{t(e)}-\omega_{o(e)}=-(E^T\omega)_e\, .
\ee
The degrees of freedom in this gauge transformation is $p-1$ and
($p-1$) scalar fields can be gauged away from the lagrangian. Therefore there are
($q-p+1$) physical, massless scalar fields.
As explained in \cite{HL}, the vector fields
absorb massive modes of the link scalar fields, and the zero-modes of the link fields
survive as physical fields.
Thus, except for the zero modes, the massive modes of link fields $\chi_e$ are nothing
but the Stueckelberg fields~\cite{stuekelberg}.

The physical massless scalar modes $\chi_{0}^{(i)}$ $(i=1,\ldots,q-p+1)$ are
orthogonal to the gauge transformation, {\it i.e.}, $\sum_{e\in
E}\chi_{0e}^{(i)}(E^T\omega)_e=0$. Thus the zero modes satisfy
$(E\chi_{0}^{(i)})_v=0$.
 Actually, the graph includes $(q-p+1)$ fundamental circuits,
that is, independent (undirected) cycles as subgraphs.
The number of fundamental circuit $n(G)$ is called 
the \textit{cyclomatic number} of the graph $G$ or
its \textit{nullity}. For the $i$-th fundamental circuit $C^{(i)}$,
\be
\chi_{0e}^{(i)}\propto\left\{
\begin{array}{ccl}
\pm 1&~&{\rm if}~e\in E(C^{(i)})\\
0&~&{\rm otherwise}
\end{array}
\right.\, ,
\ee
where the minus sign is chosen when the edge has the
opposite direction to an orientation of the fundamental circuit,
satisfies $(E\chi_{0}^{(i)})_v=0$.
Then $\chi_{0}^{(i)}$ is represented by rows of the \textit{fundamental tie set matrix}
$F_f$, since $rank \, F_f = n(G)$ 
\footnote{For the definition of fundamental circuits and fundamental tie set matrix, in
other words, fundamental circuit matrix, and for a related theorem see
\cite{graphtheory}.} .  For example,
for the graph including a (directed) fundamental circuit of length three, the incidence matrix
includes
\be
\left(
\begin{array}{rrr}
1 & 0 & -1 \\
-1 & 1 & 0 \\
0 & -1 & 1 
\end{array}\right)\, ,
\ee
as a submatrix. Then $\chi_0\propto (1, 1, 1, 0,\ldots, 0)$ is a zero mode.

Now and then the part of the lagrangian for the link fields
gives a mass term for the vector fields.
For example, we consider a cycle graph $C_5$. The $(mass)^2$ matrix for vector
fields takes the form
\be
(gv)^2\Delta(C_5)=(gv)^2\left(
\begin{array}{rrrrr}
2 & -1 & 0 &0&-1\\
-1 & 2 & -1 &0&0\\
0 & -1 & 2 &-1&0\\
0& 0 & -1 &2&-1\\
-1 & 0 &0& -1 &2
\end{array}\right)\, .
\label{cm}
\ee
Up to the dimensionful coefficient $g^2v^2$,
this matrix is identified with the Laplacian matrix for the graph $C_p$,
the cycle graph with $p$ vertices.
We find, indeed, any theory space can be associated with the graph.

Next we will manage to perform gauge fixing.
We choose the gauge fixing term as
\be
{\cal L}_{gf}=-\sum_{v\in
V}\frac{1}{2g^2\xi}(\partial_{\mu}A^{\mu}_v-\xi (gv)^2(E\chi)_v)^2\, ,
\ee
where $(E\chi)_v$ means $\sum_{e\in E}E_{ve}\chi_e$.

For this gauge choice, we introduce the ghost field and its lagrangian
\be
{\cal L}_{ghost}=\sum_{v\in V}\overline{c}_v[-(\partial^{2}+\xi(gv)^2EE^T)c]_v\, ,
\ee
because $\delta(\partial_{\mu}A^{\mu}_v-\xi (gv)^2(E\chi)_v)=\partial^2\omega_v+\xi
(gv)^2(EE^T\omega)_v$.

The gauge-fixed lagrangian ${\cal L}_{V\xi}={\cal L}_{V}+{\cal L}_{gf}+{\cal
L}_{ghost}$ becomes
\bea
{\cal
L}_{V\xi}&=&-\frac{1}{2 g^2}\sum_{v\in V}\left[(\partial_{\mu}A^{\nu}_v)^2
-\left(1-\frac{1}{\xi}\right)(\partial_{\mu}A^{\mu}_v)^2\right]+
\frac{1}{2}v^2\sum_{v\in
V} A^{\mu}_v(EE^TA_{\mu})_v\nn
& &+\frac{1}{2}\sum_{e\in
E} (\partial^\mu
X_e)^2-\frac{1}{2}\xi (gv)^2\sum_{e\in E}X_e(E^TEX)_e\nn
& &+\sum_{v\in V}\overline{c}_v[-(\partial^{2}+\xi(gv)^2EE^T)c]_v\, ,
\eea
where we rewrite the scalar fields as $X_e\equiv v\chi_e$.
The massive scalar modes are the would-be Nambu-Goldstone bosons that become a
longitudinal component of vector fields, while the massles modes are
physical massless scalars. \footnote{The massive scalar modes are associated with 
the {\it fundamental cutset matrix} $C_f$, since $F_f C_f^{T} =0$
\cite{Algebraic_graphtheory}.}

If we choose $\xi=1$ gauge, we can apparently find that vector fields, (physical
and unphysical) scalar fields and ghost fields have the same mass spectrum up to zero
modes, since $EE^T$ and $E^TE$ have the same nonzero eigen values.%
\footnote{Indeed, $\tr e^{-EE^T t}-\tr e^{-E^TE t}=p-q$ is an  `index theorem' in
graph theory as well as the theory of nonnegative matrices.}

The treatment of the `unexpected' scalars in phenomenological point of view
will be discussed after considering the coupling to fermions in the successive
section.

\section{fermions on a graph}
\label{sec:4}

We associate right-handed fermion fields with vertices of a graph $G$ and 
left-handed fermion fields with edges of $G$.
The lagrangian density for Dirac fields associated with the directed graph can be
written by
\be
{\cal L}_f=
\sum_{v\in V}\overline{\psi}_{Rv}i\gamma^\mu\partial_\mu\psi_{Rv}+
\sum_{e\in E}\overline{\psi}_{Le}i\gamma^\mu\partial_\mu\psi_{Le}-m(\sum_{e\in
E}\sum_{v\in V}\overline{\psi}_{Le}E^T_{ev}\psi_{Rv}+h.c.)\, .
\ee
The equations of motion are derived from this lagrangian as
\bea
\partial^2\psi_{Rv}+m^2(EE^T\psi_R)_v&=&0\, ,\nn
\partial^2\psi_{Le}+m^2(E^TE\psi_L)_e&=&0\, .
\eea
Here we have already known that the graph laplacian matrix $\Delta\equiv EE^T$ has a single
zero eigenvalue for a simple connected graph~\cite{Mohar}. Moreover it is well known
that the matrix
$E^TE$ has the same eigenvalues as $EE^T$ and $(q-p)$ zero modes. Therefore the
particle spectrum contains one right-handed Weyl fermion, $(q-p+1)$ left-handed Weyl
fermions (or, one massless Dirac fermion and $(q-p)$ left-handed fermion), and $(p-1)$
massive Dirac fermions.

Now we introduce the coupling between gauge and link fields.
In addition to (\ref{ga1}), we will impose the gauge symetry and
assume the following gauge transformation on fermions:
\bea
\psi_{Rv}&\rightarrow& \exp({i\omega_v})\psi_{Rv}\, ,\nn
\psi_{Le}&\rightarrow& \exp({i\omega_{o(e)}})\psi_{Le}\, .
\eea
The invariant lagrangian is
\bea
{\cal L}_f&=&
\sum_{v\in V}\overline{\psi}_{Rv}i\gamma_\mu(\partial^\mu-iA^{\mu}_v)\psi_{Rv}+
\sum_{e\in
E}\overline{\psi}_{Le}i\gamma_\mu(\partial^\mu-iA^{\mu}_{o(e)})\psi_{Le}\nn
& &-m(\sum_{e\in
E}\overline{\psi}_{L\,e}(\psi_{R\,o(e)}-U_e\psi_{R\,t(e)})+h.c.)\nn
&=&
\sum_{v\in V}\overline{\psi}_{Rv}i\gamma_\mu D^\mu_v\psi_{Rv}+
\sum_{e\in
E}\overline{\psi}_{Le}i\gamma_\mu D^\mu_e\psi_{Le}\nn
& &-m(\sum_{e\in
E}\sum_{v\in V}\overline{\psi}_{Le}\hat{E}^\dagger_{ev}\psi_{Rv}+h.c.)\, ,
\eea
where the weighted incidence matrix $\hat{E}$ is defined as
\be
(\hat{E})_{ve}=
\left\{\begin{array}{ccl}
1 &~& ~{\rm if}~v=o(e) \\
-U^\dagger_e &~& ~{\rm if}~v=t(e) \\
0 &~& ~{\rm otherwise}
\end{array}\right.
\, .
\ee
The $(mass)^2$ matrix, or modified graph Laplacian can be read as
\be
(\hat{\Delta})_{vv'}=(\hat{E}\hat{E}^\dagger)_{vv'}=
\left\{\begin{array}{ccl}
deg(v) &~& ~{\rm if}~v=v' \\
-U^\dagger_e &~& ~ e=[v',v] \in E(G)\\
-U_e &~& ~ e=[v,v'] \in E(G)\\
0 &~& ~{\rm otherwise}
\end{array}\right.
\, .
\ee


Getting the gauge field lagrangian in the previous section and the fermion
lagrangian together, we have a QED-like theory on a graph.
 When we investigate the model in  view of quantum
theory, we find that there appears chiral anomaly in general except for $p=q$ case.
The cancellation of anomaly requires more charged fermion species.
We do not treat the problem of anomaly in the present paper.
Another problem for phenomenologically viable models is the existence of exactly
massless fermions.
\footnote{Even in the $p=q$ case a massless Dirac field appears.}
Of course additional mass term can be introduced into the model,
but the origin of such a small `electron mass' is not discussed here.

Now we study the coupling between zero-mode fields, which describes the low-energy
physics $E\ll v, m$.
The lowest-order interactions can be read as
\bea
{\cal L}_{int}&=&
\sum_{v\in V}\overline{\psi}_{Rv}\gamma_\mu A^{\mu}_v\psi_{Rv}+
\sum_{e\in
E}\overline{\psi}_{Le}\gamma_\mu A^{\mu}_{o(e)}\psi_{Le}\nn
& &-\frac{m}{v}(\sum_{e\in
E}\overline{\psi}_{L\,e}iX_e\psi_{R\,t(e)}+h.c.)\, .
\eea

We consider here the simplest case, $p=q$. Then the graph contains one circuit
$\tilde{C}_{\tilde{p}}$ with length $\tilde{p}~(\leq p)$.
The zero-mode fields of $A^\mu$ and $\psi_R$ is expressed as
\be
A^\mu_v=\frac{1}{\sqrt{p}}A_0^\mu\, ,\qquad
\psi_{Rv}=\frac{1}{\sqrt{p}}\psi_{R0}\qquad  \forall v \in V(G)  \, .
\label{v0}
\ee
On the other hand, 
The zero-mode fields of $X$ and $\psi_L$ is expressed as
\be
X_{e}=\frac{1}{\sqrt{\tilde{p}}}X_{0}\, ,\qquad
\psi_{Le}=\frac{1}{\sqrt{\tilde{p}}}\psi_{L0}\qquad \forall e \in E(\tilde{C}_{\tilde{p}}) \, .
\ee
Then the zero-mode interactions can be written as
\bea
{\cal L}_{int0}&=&
\frac{1}{\sqrt{p}}\overline{\psi}_{R0}\gamma_\mu A^{\mu}_0\psi_{R0}+
\frac{1}{\sqrt{p}}\overline{\psi}_{L0}\gamma_\mu A^{\mu}_{0}\psi_{L0}\nn
& &-\frac{1}{\sqrt{p}}\frac{m}{v}(\overline{\psi}_{L\,0}iX_0\psi_{R\,0}+h.c.)\, .
\eea
The gauge coupling is $g/\sqrt{p}$, while the link scalar coupling is
$(m/v)/\sqrt{p}$, which is the same order as the gauge coupling if
the fermion mass equals to vector boson mass, $m=gv$.

We do not know the massless scalar interaction in our real world.
One way to avoid the difficulty in the existence of scalars is that  the
massless scalar is assumed to interact very weakly with matter fields.
Unfortunately, here we found that the suppression of the scalar interaction due to the
choice of a graph cannot be expected in general.  We can only arrange the two scales
$m$ and
$v$ for this purpose.

A special case is the choice of a graph with $p=q+1$, called the 
tree (graph)~\cite{graphtheory}.
All link scalar fields are absorbed by massive vector
bosons, leaving a massless vecor field. 
The simplest tree graph, path graph
$P_p$~\cite{graphtheory}, corresponds to the dimensional
deconstruction of an orbifold $S^1/Z_2$.

Another method to discard the massless scalar is incorporation of
the plaquette-type term, ${\rm Re}\,\,( U_{e_1}\cdots U_{e_{q'}})$ where
$ {e_1},\cdots, {e_{q'}}\in E(C)$ and $C$  is a cycle in a graph, in the
lagrangian.
The plaquette-type term will be studied elsewhere.


In the rest of the present paper, we will concentrate on the study of
one-loop UV divergence of the model.

\section{One-loop divergences in effective action}
\label{sec:5}

We investigate the one-loop divergence in the model by calculating the effective
action by heat-kernel method with background fields~\cite{Vass}. In this section, we
use the Euclidean signature for the metric. The gamma matrices satisfy
$\{\gamma^{\mu},\gamma^{\nu}\}=-2\delta^{\mu\nu}$. We assemble the fermion fields as
\be
\Psi\equiv\left(\begin{array}{c}\psi_R\\ \psi_L\end{array}\right)\, ,
\ee
and define a derivative operator
\be
i{\cal D}\equiv\left(\begin{array}{cc}
iD_V & -m\hat{E}\\
-m\hat{E}^\dagger & iD_E
\end{array}\right)\, ,
\ee
and
\be
i{\cal D}^\dagger\equiv\left(\begin{array}{cc}
iD_V & ~m\hat{E}~\\
~m\hat{E}^\dagger ~ & iD_E
\end{array}\right)\, ,
\ee
where
$D_V=diag(\gamma^{\mu}D^{\mu}_{v_1},\gamma^{\mu}D^{\mu}_{v_2},
\ldots,\gamma^{\mu}D^{\mu}_{v_p})$ and
$D_E=diag(\gamma^{\mu}D^{\mu}_{e_1},\gamma^{\mu}D^{\mu}_{e_2},
\ldots,\gamma^{\mu}D^{\mu}_{e_q})$.
Then the Euclidean lagrangian is expressed as ${\cal L}_f=\Psi^\dagger i{\cal D}\Psi$.

As a preparation, we write the quatratic operator ${\cal D}^\dagger{\cal D}$
explicitly as
\be
{\cal D}^\dagger{\cal D}=\left(\begin{array}{cc}
D_V^2+m^2\hat{E}\hat{E}^\dagger & im(D_V\hat{E}-\hat{E}D_E)\\
im(D_E\hat{E}^\dagger-\hat{E}^\dagger D_V) & D_E^2+m^2\hat{E}^\dagger\hat{E}
\end{array}\right)\, ,
\ee
where
\be
(D_V^2)_{vv'}=\left\{
\begin{array}{ccl}
-(D_v^{\mu})^2-\frac{i}{2}\gamma^{\mu}\gamma^{\nu}F^{\mu\nu}_v&~&{\rm if}~v=v'\\
0&~&{\rm otherwise}
\end{array}
\right.\, ,
\ee
\be
(D_E^2)_{ee'}=\left\{
\begin{array}{ccl}
-(D_e^{\mu})^2-\frac{i}{2}\gamma^{\mu}\gamma^{\nu}F^{\mu\nu}_{o(e)}&~&{\rm if}~e=e'\\
0&~&{\rm otherwise}
\end{array}
\right.\, ,
\ee
\be
(D_V\hat{E}-\hat{E}D_E)_{ve}=\left\{
\begin{array}{ccl}
-\gamma^{\mu}(D^{\mu}U_e)^{\dagger}&~&{\rm
if}~v=t(e)\\ 0&~&{\rm otherwise}
\end{array}
\right.\, ,
\ee
and
\be
(D_E\hat{E}^\dagger-\hat{E}^\dagger D_V)_{ev}=\left\{
\begin{array}{ccl}
-\gamma^{\mu}(D^{\mu}U_e)&~&{\rm
if}~v=t(e)\\ 0&~&{\rm otherwise}
\end{array}
\right.\, .
\ee

The effective action at one-loop level can be written as~%
\footnote{See \cite{Vass} and references therein.}
\be
\Gamma=-\frac{1}{2}\Tr\ln ({\cal D}^\dagger{\cal D})=
\frac{1}{2}\int_0^{\infty}\frac{dt}{t}
\Tr\, e^{-{\cal D}^\dagger{\cal D}t}=
\frac{1}{2}\int_0^{\infty}\frac{dt}{t}
\int d^4x\,\tr\langle x|e^{-{\cal D}^\dagger{\cal D}t}|x\rangle\, .
\ee
Here we use
\be
\langle x|f(D_{\mu})|x\rangle=\int\frac{d^4k}{(2\pi)^4}
\langle x|f(D_{\mu}+ik_{\mu})|0\rangle\, ,
\ee
where $|0\rangle$ is a zero momentum state ($\langle x|0\rangle=1$),
to calculate the effective lagrangian~\cite{Vass}.

To evaluate the $t^2$ term in the expansion of the integrand, we need the following
explicit form
\be
\tr
(D_V^2)^2=2\sum_{v\in V}\left[(D_v^{\mu})^2(D_v^{\nu})^2+
\frac{1}{2}F^{\mu\nu}_vF^{\mu\nu}_v\right]\, ,
\ee
and
\be
\tr
(D_E^2)^2=2\sum_{e\in E}\left[(D_e^{\mu})^2(D_e^{\nu})^2+
\frac{1}{2}F^{\mu\nu}_{o(e)}F^{\mu\nu}_{o(e)}\right]\, ,
\ee
where the coefficient two comes from the projection to the left/right handed
fermions.

Therefore logarithmically divergent part (with the Euclidean signature) turns out to
be
\be
\frac{1}{24\pi^2}\int\frac{dt}{t}\left[
\sum_{v\in V}\frac{1}{4}(F_v^{\mu\nu})^2+
\sum_{e\in E}\frac{1}{4}(F_{o(e)}^{\mu\nu})^2+
\frac{3}{2}m^2\sum_{e\in E}{|D^{\mu} U_e|}^2+\cdots\right]\, .
\ee
The coefficients of kinetic terms of gauge and link fields are logarithmically
divergent.

In particular, the gauge coupling runs logarithmically. If every vertex is an origin
of an edge for $p=q$ graph, the beta function is the same as usual QED:
\be
\mu\frac{dg}{d\mu}=\frac{g^3}{12\pi^2}\, .
\ee
In general case, if we define the individual gauge coupling for each
gauge field on a vertex, 
such that the kinetic term becomes $\sum_{v\in
V}\frac{1}{4g_v^2}F_{\mu\nu~v}F^{\mu\nu}_v$, their beta functions are
\be
\mu\frac{dg_v}{d\mu}=\frac{(1+d^+(v))g_v^3}{24\pi^2}\, ,
\ee
where $d^+(v)$ is the {\it outgoing degree} of a vertex $v$, which is the number of edges whose origin is $v$.
\footnote{Note that the average of $d^+(v)$ is $q/p$.}

The running of the individual gauge coupling is interesting for the possibility of
variating the mass spectrum. However, the existence of zero modes is still unchanged,
expressed as in (\ref{v0}).
Thus this running-coupling effect is not useful for symmetry breaking even when the
model is generalized to non-Abelian one.

For non-Abelian case, the quantum fluctuation of the gauge fields and link fields
also induce the logarithmic divergent contribution to the gauge and link field
kinetic terms.

\section{One-loop finiteness in effective potential}
\label{sec:6}

The effective potential for constant background link field can be written as
\bea
V&=&-\frac{1}{2}
\int_0^{\infty}\frac{dt}{t}\int\frac{d^4k}{(2\pi)^4}e^{-k^2t}
2\left[\tr_p\exp(-m^2\hat{E}\hat{E}^\dagger t)+\tr_q\exp(-m^2\hat{E}^\dagger\hat{E}
t)\right]\nn
&=&-\frac{1}{(4\pi)^2}
\int_0^{\infty}\frac{dt}{t^3}
\left[\tr_p\exp(-m^2\hat{E}\hat{E}^\dagger t)+\tr_q\exp(-m^2\hat{E}^\dagger\hat{E}
t)\right]\, ,
\label{ep}
\eea
where $\tr_M$ means the trace only on $(M\times M)$ matrices.

Since $\tr_p (\hat{E}\hat{E}^\dagger)^n=\tr_q(\hat{E}^\dagger\hat{E})^n$ for
$n\ge 1$, the integrand can be expanded as
\be
\frac{1}{2}\left[\tr_p\exp(-m^2\hat{E}\hat{E}^\dagger
t)+\tr_q\exp(-m^2\hat{E}^\dagger\hat{E} t)\right]=\frac{p+q}{2}-m^2\tr_p\hat{\Delta}
t+\frac{1}{2}m^4\tr_p\hat{\Delta}^2t^2+O(t^3)\, .
\ee

Owing to $U_eU^\dagger_e=1$, $\tr_p\hat{\Delta}=\tr_pD$ obviously and
$\tr_p\hat{\Delta}^2=\tr_pD^2+\tr_pD$. In other words, the same relation as that on $\Delta$
holds.
The expression (\ref{ep}) includes divergences, but they does not depend on the backgroud link fields.

This is the origin of the one-loop finiteness of the scalar potential in the
deconstructed theory. This nature is preserved for non-Abelian generalization.

In the present Abelian case, the zero-mode field of the link variable acquires mass by
the one-loop quantum effect. 
The explicit calculation for models with a cycle graph $C_N$
can be carried out as
in~\cite{HL,KSS}.

\section{Graph Hosotani Mechanism?}
\label{sec:7}

If some non-Abelian gauge symmetry is introduced, a symmetry breaking mechanism
becomes possible in the theory on a graph, just as in the case of the Hosotani
mechanism~\cite{Hosotani}.

Let the length of the shortest cycle $c=(e_1,\ldots e_N)$ in $G$ be $N
(\ge 3)$. The trace of the kernel of $\hat{\Delta}$, $\tr\exp(-\hat{\Delta}t)$, for
matter fields coupled to the link fields includes a term ${\rm Re}\,\,\tr \,
U_{e_1}\cdots U_{e_N}$ and its coefficient is
$O(t^N)$. Therefore the one-loop effective potential for the zero mode 
of link fields is finite (up to a field-independent divergence).

If we take a graph $C_n$ into a model and consider the limit $n\rightarrow\infty$, 
the model reduces to the original Hosotani model~\cite{Hosotani}
(it should be read $A_5\sim \chi,$ where $U=e^{-i\chi}$).

The realization of non-Abelian symmetry breaking in the field theory on a graph may
not be so easy as in the Hosotani models.
The simplest idea is that we abandon the {\it local symmetry on a whole graph}.
Suppose that we dare to replace such a term 
$\overline{\psi}_{L\,e}(\psi_{R\,o(e)}-U_e\psi_{R\,t(e)})$ by
$\overline{\psi}_{L\,e}(\psi_{R\,o(e)}-\psi_{R\,t(e)})$ on some edges.
Then the {\it local} symmetry on a graph disappears but {\it global} symmetry on a
graph such as $\psi_{Rv}\rightarrow \exp({i\omega})\psi_{Rv}$,
$\psi_{Le}\rightarrow \exp({i\omega})\psi_{Le}
$ (where $\omega_v(x)=\omega(x)$ for all $v\in V$ and $\omega_e(x)=\omega(x)$ for all
$e\in E$) remains. Thus the model still has usual local gauge symmetry in spacetime,
at classical level.  In this case, the induced term like ${\rm Re}\,\tr \,
U_{e_1}\cdots U_{e_N}$, where some $U$s are replaced by unity, may lead to a novel
phase structure of vacuum.

We will study the dynamical symmetry breaking in field theory on a graph which has
edges with and without some weight functions elsewhere.

\section{summary and prospects}
\label{sec:f}

In conclusion, we clarified the divergences of the one-loop effective lagrangian in
the Abelian gauge field theory on graphs. 

We must consider the following possibilities.
To consider the generalization of the Hosotani model,
we should investigate 
 non-Abelian gauge theory on a graph.
We also need 
adjoint matter fields or fields in other representations on vertices or edges.
At the same time,
we should study the possible inclusion of plaquette-like self-interaction of link
fields in bare lagrangian.
By the way, to consider superfields on a graph is also an interesting subject.

We are interested also in the two-loop effective action.
We still hope that the knowledge of algebraic graph theory may be
useful to investigate higher-loop divergence as well as tree-level calculation of
reaction amplitude mediated by the excited modes.

\begin{acknowledgments}
We would like to thank Y.~Cho for his valuable comments
and for reading the manuscript.
\end{acknowledgments}



\end{document}